\documentclass[prl,twocolumn,showpacs,nobibnotes,floatfix,superscriptaddress]{revtex4}

\usepackage{graphicx,eucal}

\begin{document}

\title{Efficient Construction of Photonic Quantum Computational Clusters}
\author{Gerald Gilbert, Michael Hamrick and Yaakov S. Weinstein\\
\small \it Quantum Information Science Group\\
{\sc  Mitre} \\
\small \it 260 Industrial Way West, Eatontown, NJ 07724 USA \\
\tt{E-mail: \{ggilbert, mhamrick, weinstein\}@mitre.org}}

\begin{abstract}
We demonstrate a method of creating photonic two-dimensional cluster states that is 
considerably more efficient than previously proposed approaches. Our method uses
only local unitaries and type-I fusion operations. The increased efficiency of our method compared to previously proposed constructions is obtained by identifying and exploiting local equivalence properties inherent in cluster states.
\end{abstract}


\maketitle

Cluster states are entangled states constructed in such a way as to enable universal quantum computation (QC), effected solely by suitable measurements performed on the constituents of the cluster~\cite{BR1,BR2,BR3}. In this paper we present a new method for constructing {\em photonic} clusters that is more efficient than previously presented methods. 

Photonic quantum computation has received significant attention in recent years. The initial work on photonic QC considered different {\em circuit-based} approaches~\cite{fn1}. Since quantum computational logic transformations, such as the CNOT gate, require a mechanism by which qubits directly interact, the choice of photons as qubits in turn motivated the choice of nonlinear Kerr-type media in the first analyses of photonic QC~\cite{THL}. Although this approach in principle enables photonic entangling gates, the practical difficulties associated with the use of Kerr-type media made this method appear problematic.

Interest in photonic QC was renewed with the appearance of the work of Knill, Laflamme and Milburn (KLM) in 2001~\cite{KLM}. This approach makes use of linear optics, combined with measurements carried out on ancillary photons, in order to circumvent the difficulties associated with the use of nonlinear media. Although it avoids the use of nonlinear media, the KLM approach to linear optics quantum computation (LOQC) is nevertheless problematic due to the inefficiency associated with the necessity of dealing with extremely large numbers of ancillary photons~\cite{N}.

Both of these approaches to photonic QC, the nonlinear approach and the linear approach, are formulated within the circuit-based paradigm. With the discovery of the cluster-based paradigm, it became natural to explore the possibility of using photons as the nodes in a cluster.

Nielsen noted \cite{N} that a photonic cluster could furnish a more efficient realization of a quantum computation than a photonic circuit if certain techniques from LOQC were used to {\em build} the photonic cluster (as opposed to directly {\em executing} the computation itself). Browne and Rudolph \cite{BR} refined this idea, and presented a more efficient scheme for the construction of photonic clusters. In this scheme, the suggestion of Nielsen to use LOQC-derived operations to construct the cluster, is replaced by a proposal to use simpler ``fusion" operations to construct the cluster. A number of additional methods for constructing clusters have been suggested \cite{DR,CCWD} and small photonic cluster states have been experimentally implemented \cite{Zeil,K,Zhang}.

In this paper we present a method for constructing generic clusters that, in the case of photonic clusters in particular, is considerably more efficient than previously proposed methods. The increased efficiency is achieved by identifying certain equivalence classes of cluster configurations, which are exploited to reduce the number of distinct transformations that are needed to generate clusters suitable for carrying out universal QC. Our technique makes use of the properties of equivalent graph states under local unitaries and graph isomorphisms, and combines this with the use of type-I fusion operations. We thus entirely avoid the use of resource-costly type-II fusion operations. Our approach results in a significant increase in cluster construction efficiency. In particular, we show that our method is more efficient, in terms of resources used, than that of \cite{BR}.

Consider the following equivalence, described in \cite{HEB,Zeil}. A 4-qubit linear cluster, or chain, is equivalent to a $2\times 2$ box cluster up to Hadamard rotations and a swap operation. We denote the action of a Hadamard rotation on the $j$th qubit by the symbol $H_j$. The symbol $SWAP_{j,k}$ denotes the swap operation acting on qubits $j$ and $k$, which can be realized by simply relabeling the qubits. Applying $H_2\otimes H_3$ to a 4-qubit cluster chain and exchanging the labels of qubits 2 and 3 effectively adds a bond between qubits 1 and 4. This is illustrated in Figure 1a.

Note that Figure 1a depicts a transformation that involves {\em only the four qubits shown} in the diagram. We extend this transformation by embedding the
4-qubit chain in a larger linear structure of arbitrary size, as depicted in Figure 1b. Here the chain extends arbitrarily far in both directions. The presence of the extensions in both directions of the initial chain in Figure 1b reflects the existence of additional entanglement correlations between qubits 1 and 4 and the qubits along the chain extensions. 
One can show that these additional entanglement correlations are in fact preserved by the same transformation utilized in Figure 1a~\cite{HEB}.

We have thus extended the ``box construction" method given in \cite{Zeil} to obtain a ``box-on-a-chain construction" method. We now apply our ``box-on-a-chain construction" method to the problem of generating generic clusters suitable for carrying out universal QC. Following the suggestion in references~\cite{BR} and \cite{N}, we adopt the use of an $L$-shaped lattice, constructed from arbitrary length chains, as a basic ``building block" with which to construct clusters capable of carrying out universal QC. In this way the $L$-shaped lattice serves as a standard figure-of-merit with which to measure the efficiency of the construction of general quantum computational clusters.

Our efficient method of constructing the $L$-shape is illustrated in Figure 1c. We start with a cluster chain, drawn suggestively as the left-most element in Figure 1c. After carrying out the transformation already described to obtain the middle element in Figure 1c, one then applies a $\sigma_z$ measurement on qubit 2 to produce the $L$-shape in the right-most element in Figure 1c. This operation deletes 2 bonds from the cluster chain due to the measurement of qubit 2. Note that this technique requires no probabilistic operations, {\em and carries a net cost of only 2 cluster chain bonds}. In contrast, Browne and Rudolph use the probabilistic type-II fusion operation to build the $L$-shape requiring, on average, 8 bonds from previously constructed cluster chains~\cite{BR}.

\begin{figure}
\includegraphics[height=5.8cm]{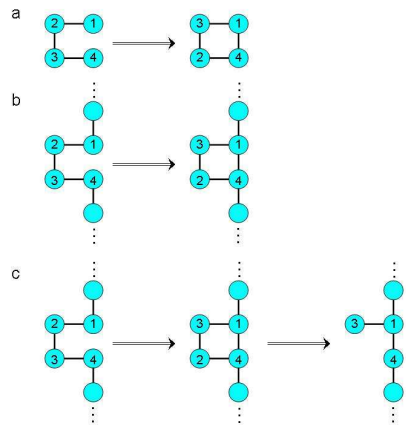}
\caption{\label{f1}
(Color online) Figure 1a shows how Hadamard and swap operations are used to transform a chain into a box shape. In Figure 1b the above operations are used to prepare the ``box-on-a-chain" cluster state. In Figure 1c, the ``box-on-a-chain" form is used to create the basic $L$-shape used as a building block for cluster states capable of universal quantum computation. To achieve this, a $\sigma_z$ measurement on qubit 2 deletes all bonds involving the measured qubit, deterministically giving the desired $L$-shape cluster at a cost of only two bonds.}
\end{figure}

Our technique for constructing the basic $L$-shape cluster ``building blocks" is generic, and not specific to photonic quantum bits. In order to yield a complete, integrated method of efficiently constructing general {\em photonic} clusters capable of universal QC, we can combine our method of constructing $L$-shapes with the already-known type-I fusion operation of Browne and Rudolph. This integrated cluster generation method is significantly more efficient than previously proposed approaches: the increased efficiency derives from the improved efficiency of our $L$-shape generation technique as compared to the costly technique based on the use of type-II fusion operations. Type-II fusion operations are not needed in our approach. 

\begin{figure}
\includegraphics[height=5.8cm]{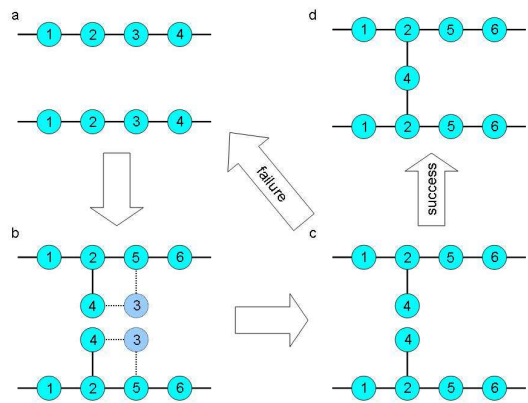}
\caption{\label{f4}
(Color online) Two $L$-shape clusters, of the type constructed in Figure \protect\ref{f1}c, are linked {\em via} a type-I fusion operation. Begin with separate chains as in Figure 2a. Figures 2b and 2c illustrate the deterministic transformation of the chains into 2 $L$-shapes. A type-I fusion operation is applied: if successful, the sideways $H$-shape shown in Figure 2d is created; if unsuccessful, recover two separate chains, return to Figure 2a and repeat.}
\end{figure}

The general method for constructing a 2-dimensional cluster, such as the ``sideways $H$-shape" illustrated in Figure 2d, proceeds as follows. We start with the assumption that we have two chains of arbitrary length, as illustrated in Figure 2a. Given two initial chains, we {\em deterministically} transform each of them into an $L$-shape using our method described above, shown in Figure 2b, at a cost of 2 bonds for each $L$-shape, or a net cost of 4 bonds. (In constrast, transforming two chains into 2 $L$-shapes, using the {\em non-deterministic} type-II fusion operations of Browne and Rudolph, would cost 8 bonds on average for each $L$-shape, and would thus result in an average net cost of 16 bonds.) Then, making use of the type-I fusion operation of Browne and Rudolph, we attempt to adjoin the 2 completed $L$-shapes illustrated in Figure 2c to form the desired 2-dimensional $H$-shape cluster. Since the type-I fusion operation succeeds with probability $1/2$, there are two possible outcomes: (1) If the joining operation {\em succeeds} as shown in Figure 2d we are done, and the desired 2-dimensional cluster has been built at a net cost of 4 bonds. In contrast, had the initial $L$-shapes been formed using type-II fusion operations, the average net  cost of the 2-dimensional cluster would have been 16 bonds. (2) If the joining operation {\em fails}, each $L$-shape reverts to a chain, returning us to the state illustrated in Figure 2a, incurring a net additional cost of 2 bonds. Then, the process described above is iterated, beginning with the new chains that resulted from the failure of the attempted 
joining operation, until success is achieved. Upon such iteration, the average net cost for a successful outcome is 10 bonds if the initial (and subsequent) chains are transformed into $L$-shapes using our technique, but the average net cost would have been 34 bonds had the various chains been transformed into $L$-shapes using the type-II fusion operation of Browne and Rudolph.

We have presented a method for constructing 2-dimensional photonic clusters by making use of the type-I fusion operation of Browne and Rudolph combined with our technique for transforming a given cluster chain into the basic $L$-shape. As demonstrated above, this method of generating a 2-dimensional $H$-shape cluster is significantly more efficient than the original proposal of \cite{BR} that makes use of type-I {\em and} type-II fusion operations. We note that $H$-shapes are comprised of a single ``rung" connecting two chains. These $H$-shapes can be grown into ``sideways ladder" shapes that possess additional rungs, by repeatedly applying our method along the length of an initial $H$-shape. In addition, two-dimensional clusters with greater depth than an $H$-shape (or a sideways ladder) can be built by adjoining parallel chains to a given $H$-shaped cluster. This is done, one depth level at a time, by applying our method to a given additional chain and either of the ``outer" sides of the starting $H$-shape. In this way, making use of our technique for creating basic $L$-shapes out of chains, a 2-dimensional cluster of any complexity can be formed, with a significant increase in efficiency compared to previous approaches~\footnote{More general $H$-shapes (and more general ``sideways ladder" shapes), in which the rungs connecting adjacent chains include more than one qubit node, are constructed by adjoining modified $L$-shapes that have been augmented using type-I fusion operations to incorporate additional qubit nodes. In addition, one constructs simple bonds connecting adjacent chains ({\em i.e.}, ``nodeless rungs"), by removing nodes as required using $\sigma_y$ measurements.}.

Our cluster construction method employs only local unitary rotations and type-I fusion operations: type-II fusion operations are neither needed nor used, which results in a significant increase in the efficiency of photonic cluster construction as discussed above. Although our method makes no use of type-II fusion operations, this does not compromise the generality or diminish the flexibility of the method. As an illustration of this flexibility, we now discuss a sample of typical cluster shapes that can be constructed making use only of local unitaries and type-I fusion operations.

\begin{figure}
\includegraphics[height=5.8cm]{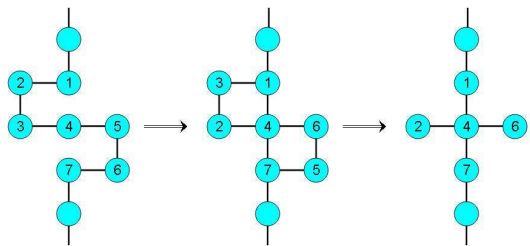}
\caption{\label{f2}
(Color online) Starting with a cluster chain, two bonds are added 
via local unitaries. $\sigma_z$ measurements on qubits 3 and 5 (at a 
cost of four bonds) yield the desired cluster state cross.}
\end{figure}

\begin{figure}
\includegraphics[height=5.8cm]{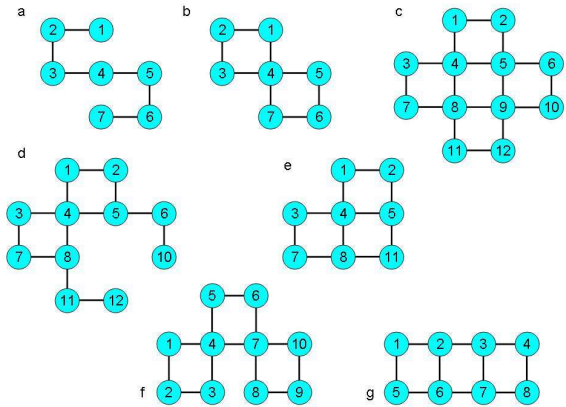}
\caption{\label{f3}
(Color online) Representative cluster shapes illustrating the generality and flexibility of the efficient cluster construction method described in the text.}
\end{figure}

As a first example, we note that a cluster chain can be deterministically transformed into a cross shape. Following the progression illustrated in Figure \ref{f2}, apply $H_2\otimes H_3\otimes H_5\otimes H_6$ followed by $SWAP_{2,3}\circ SWAP_{5,6}$. This forms bonds between qubits 1 and 4, and between qubits 4 and 7. Subsequent execution of $\sigma_z$ measurements on qubits 3 and 5 creates the desired cross shape at a cost of only 4 bonds and involves no probabilistic operations.
Continuing with examples of constructions of additional shapes, we begin with the 7-qubit chain suggestively drawn in Figure 4a. Noting the similarity of Figure 4a with the left-most element of Figure 3, we may apply the same Hadamard and swap operations used in Figure 3 in order to obtain the structure shown in Figure 4b. The cluster shape in Figure 4b can then be used as an alternative to the previously discussed $L$-shape as a basic building block to construct general two-dimensional clusters. To illustrate the use of this cluster shape as a building block for more general shapes, we note that two cluster shapes of the type presented in Figure 4b, when adjoined by the successful application in succession of two type-I fusion operations, yield the cluster depicted in Figure 4c. If the second of the two attempted type-I fusion operations used to build the cluster of Figure 4c fails, the cluster that remains is illustrated in Figure 4d~\footnote{If the first of the two attempted type-I fusion operations fails, the clusters that result may be salvaged by applying to them another type-I fusion operation, and two $\sigma_z$ measurements. If {\em this} type-I fusion succeeds, an additional copy of the structure illustrated in Figure 4b will be produced. This may then be used as a resource to continue building the overall desired cluster structure.}. This shape in turn may be used to attempt to build the cluster illustrated in Figure 4e. One proceeds by using type-I fusion operations to fuse qubits 10 and 12, followed by a $\sigma_y$ measurement applied to qubits 6 and 11. If the fusion fails, rather than performing the $\sigma_y$ measurement, we attempt to fuse qubits 6 and 11 {\em via} type-I fusion. If that also fails one recovers the basic structure of Figure \ref{f3}b. If either fusion operation succeeds ({\em i.e.}, applied to the pair \{10,12\} or \{6,11\}), we continue building a more general structure making use of the cluster of Figure \ref{f3}e. Further exploring our approach to generating basic structures, we note that the cluster shapes in Figures 4f and 4g are generically useful. The cluster shape illustrated in Figure 4f may be deterministically constructed by starting with a 10-qubit chain and applying suitable Hadamard and swap operations by analogy with the transformation of a 7-qubit chain into the shape given in Figure 4b. We obtain the useful structure depicted in Figure 4g as follows. Beginning with a 9-qubit chain, we join the ends using type-I fusion to obtain an 8-qubit ring. If this operation succeeds, we apply Hadamard operations on qubits 1,4,5 and 8, and $SWAP_{1,5}SWAP_{4,8}$. In practice, one continuously builds clusters such as those illustrated in Figures \ref{f3}c, \ref{f3}e and \ref{f3}g offline and keeps only the results of successful attempts (Figure \ref{f3}f is deterministically constructed and is always successfully formed). Those clusters may then be fused together (with successful type-I fusion operations) to form larger clusters, and in this way one builds arbitrary two-dimensional clusters suitable for quantum computation~\footnote{We consider the efficient construction of additional useful cluster architectures, as well as the efficient construction of the necessary initial cluster chains, in a separate paper.}.

In conclusion we have introduced and demonstrated a method of constructing two-dimensional photonic cluster states for quantum computation that is significantly more efficient than previously proposed approaches. Our technique exploits the properties of equivalent graph states under local unitaries and graph isomorphisms, and combines this with the use of type-I fusion operations. Our method entirely avoids the use of resource-costly type-II fusion operations. We demonstrated the generality and flexibility of our approach with a number of explicit representative sample constructions of useful cluster shapes.

\vspace{.6in}

\section{Acknowledgements} This research was supported under MITRE Technology Program Grant 07MSR205.

\end{document}